\documentclass[a4paper,aps,prl,floatfix,twocolumn,showpacs,nofootinbib,superscriptaddress]{revtex4-1}
\usepackage{amssymb}
\usepackage{amsmath}
\usepackage{amsfonts}
\usepackage{epsfig}
\usepackage{graphicx}
\usepackage{color}
\usepackage[latin1]{inputenc}
\usepackage{hyperref}

\begin{document}

\author{Guillem Mosquera-Do\~nate.}
\affiliation{Departament de F\'isica Fonamental, Universitat de Barcelona, Mart\'i i Franqu\`es 1, 08028 Barcelona, Spain}
\author{Mari\'an Bogu\~{n}\'a}
\affiliation{Departament de F\'isica Fonamental, Universitat de Barcelona, 
Mart\'i i Franqu\`es 1, 08028 Barcelona, Spain}
%  \email{MBmail@ub.edu}
\date{\today}

\title{Follow the Leader: Herding Behavior in Heterogeneous Populations}

\begin{abstract}
Here we study the emergence of spontaneous leadership in large populations. In standard models of opinion dynamics, herding behavior is only obeyed at the local scale due to the interaction of single agents with their neighbors; while at the global scale, such models are governed by purely diffusive processes. Surprisingly, in this paper we show that the combination of a strong separation of time scales within the population and a hierarchical organization of the influences of some agents on the others induces a phase transition between a purely diffusive phase, as in the standard case, and a herding phase where a fraction of the agents self-organize and lead the global opinion of the whole population.

\end{abstract}


\maketitle

Humans are unpredictable but collective human behavior can be predicted. Although apparently contradictory, this statement is the main hypothesis of sociophysics~\cite{Arnopoulos:1993,Galam:2012}. By analogy with (non-equilibrium) thermodynamics, the hope is that collective social phenomena can be treated as 
emerging properties of systems of interacting agents that depend on a few fundamental features of the microscopic interaction laws, rather than on the idiosyncratic character of single individuals. This hope has encouraged the study of social dynamics using the tools and models that statistical physics has been developing for the last fifty years or so~\cite{Castellano:2008ge}. Opinion dynamics is one of the better examples 
of the application of this approach. The aim here is to understand the opinion of a population of agents and the rules that govern transitions between different opinion states as a response to social influence, or the tendency of people to become like those they have social contact with~\cite{Festinger:1950}. 

Models of opinion dynamics typically show consensus states, where the dynamics is frozen. In many cases, as in the voter~\cite{Clifford:1973zc,Holley:1975} or Sznajd models, the (weighted) ensemble average opinion of the population is a conserved quantity. In such cases, the dynamics of the stochastic average opinion is governed by a purely (non-homogeneous) diffusive process without any drift, which eventually leads the system to one of the possible consensus states. It is therefore difficult to imagine how leadership can emerge in this context. In this paper, we show that leadership can, in fact, arise spontaneously in a subset of the population when there is strong heterogeneity in the time scales of the agents coupled with a hierarchical organization of their influence. Heterogeneity of time scales is present, for instance, in speculative markets, where noise traders who operate at the scale of minutes or hours coexist with fundamentalists, who operate at the scale of weeks or months. Interestingly, we reveal a pitchfork bifurcation that separates a purely diffusive phase from a phase where the most active agents lead the global state of the entire population. Our results could shed light on the dynamics of financial crises and other extreme events caused by humans.

The voter model was first introduced in 1973 to model competition between species~\cite{Clifford:1973zc,Holley:1975}. Ever since, it has been one of the most paradigmatic and popular models of opinion dynamics. Its simplicity, analytical tractability and versatility when it comes to introducing new mechanisms make it the perfect model for studying many different phenomena in both the natural and social sciences, from catalytic reaction models~\cite{EVANS:1991kx,Evans:1993fk} to the evolution of bilingualism~\cite{Castello:2006ik} or US presidential elections~\cite{Fernandez-Gracia:2014uq}. In its simplest version, the voter model is defined as follows. There is a set of $N$ interacting agents, each endowed with a binary state of opinion (sell or buy, Democrat or Republican, Windows or Mac, etc). For each time step of the simulation, an agent is randomly chosen to interact with one of their social contacts, after which that agent copies the opinion of their contact.

Heterogeneity can be introduced within the population through the activity rate of agents~\cite{Masuda:2010uq,Fernandez-Gracia:2011kx}. We assume that agents are given intrinsic activity rates $\{ \lambda_i\}$, which control the frequency at which the agents interact with their social contacts and, possibly, change their opinion. In numerical simulations, this is equivalent to choosing the next active agent, say agent $i$, with probability proportional to $\lambda_i$. The influence of one agent on others can be modeled by the probability $\mathrm{Prob}(j|i)$ that agent $i$ copies the opinion of agent $j$ when $i$ is activated at rate $\lambda_i$. When contacts take place according to a fixed social contact graph with an adjacency matrix $a_{ij}$, this probability is given by $\mathrm{Prob}(j|i)=a_{ij}/k_i$, where $k_i$ is the degree of agent $i$~\cite{Sood:2005fk,Suchecki:2005fk,Vazquez:2008je,Sood:2008kk}. If a fully connected graph pertains (equivalent to a mean-field description), this probability is simply $\mathrm{Prob}(j|i)=1/(N-1)$ for $j \ne i$ and zero otherwise. 

The dynamics of the state of the system can be described using a set of $N$ dichotomous stochastic processes $\{n_i(t)\}$ that take the value $0$ or $1$ depending on the opinion state of each agent at time $t$. If we assume that all temporal processes follow Poisson statistics, the stochastic evolution of $n_i(t)$ after an increment of time $dt$ satisfies the stochastic equation~\cite{Catanzaro:2005fk,Boguna:2009pi}:
\begin{equation}
	n_{i}(t+dt)=n_{i}(t) \left[1-\xi_{i}(t)\right] + \eta_{i}(t) \xi_{i}(t),
	\label{eq:dinamica}
\end{equation} 
where $\xi_{i}(t)$ is a random dichotomous variable that takes values:
\begin{equation}
   \xi_{i}(t) = \left\{
	\begin{array}{ll}
       1 & \mbox{with probability } \lambda_{i} dt\\
       0 & \mbox{with probability } 1-\lambda_{i} dt
     \end{array}
   \right.
\end{equation}
Notice that $\xi_i(t)$ controls whether node $i$ is activated during the time interval $(t,t+dt)$. If it is, the opinion of a neighbor will be changed according to $\mathrm{Prob}(j|i)$, so that:
\begin{equation}
   \eta_{i}(t) = \left\{
	\begin{array}{ll}
       1 &  \mbox{with probability } \displaystyle{\sum_{j=1}^N}\mathrm{Prob}(j|i) n_{j}(t) \\
       0 &  \mbox{with probability } 1-\displaystyle{\sum_{j=1}^N}\mathrm{Prob}(j|i) n_{j}(t)
     \end{array}
   \right.
\end{equation}
In principle, $\eta_{i}(t)$ should be realized only when $\xi_i(t)=1$. However, due to the particular form of Eq.~\eqref{eq:dinamica}, the value of $\eta_i(t)$ is only relevant when $\xi_i(t)=1$. Therefore, we can safely consider $\xi_i(t)$ and $\eta_i(t)$ as statistically independent random variables. 

Equation~\eqref{eq:dinamica}, supplemented with the definitions of variables $\xi_i(t)$ and $\eta_i(t)$, represents the exact stochastic evolution of the system. For instance, the ensemble average of the opinion of agent $i$, $\rho_i(t) \equiv \langle n_i(t)\rangle$ can be evaluated by taking the average of Eq.~\eqref{eq:dinamica} first over the variables $\xi_i(t)$ and $\eta_i(t)$, and then over the ensemble. This program leads to the exact differential equation:
\begin{equation}
	\frac{d\rho_{i}}{dt}=\lambda_{i}\left[\sum_{j=1}^N \mathrm{Prob}(j|i) \rho_{j}-\rho_{i}\right].
    \label{eq:evolucio}
\end{equation} 
This equation implies the existence of a global conserved magnitude~\cite{Suchecki:2005uq,Serrano:2009kx} related to the eigenvector $\phi(i)$ of eigenvalue 1 of $\mathrm{Prob}(j|i)$; that is, the solution of the equation $\sum_i \phi(i)\mathrm{Prob}(j|i)=\phi(j)$. Indeed, by multiplying Eq.~\eqref{eq:evolucio} by $\phi(i)/\lambda_i$ and summing over all agents, the right-hand side of the equation vanishes. Therefore, the weighted ensemble average of the population:
\begin{equation}
	\Phi \equiv \sum_{i=1}^N \frac{\phi(i)}{\lambda_{i}} \rho_{i}(t)=\sum_{i=1}^N \frac{\phi(i)}{\lambda_{i}} \rho_{i}(0)
	\label{eq:conservacio}
\end{equation} 
is conserved by the dynamics and thus it is a function only of the initial conditions; as we show 
above in the right-hand hand side of Eq.~\eqref{eq:conservacio}. This fact can be used to evaluate the probability of the final fate of a realization of the dynamics. For instance, the probability of ending up absorbed in the ``$1$'' consensus state is given by just $\Phi/\sum_i \phi(i)/\lambda_i$.

The results presented so far are valid for an arbitrary distribution of individual rates $\lambda_i$. However, the behavior of the system can be very different when there is a strong separation of time scales present in the system (as we say above, as in speculative markets with noise traders and fundamentalists). To shed light on this problem, hereafter we analyze a simple model with a population segregated into two groups: a fast group of size $N_f$, operating at rate $\lambda_f$; and a slow one of size $N_s$, operating at rate $\lambda_s$, with $\lambda_f> \lambda_s$. Aside from heterogeneity in their time scales, agents in a real population are also heterogeneous in terms of their influence on others. To model this effect, we assume that the probability of agent $i$ copying the opinion of agent $j$ is a function of the rate at which agent $j$ operates, 
that is:
\begin{equation}
	\mathrm{Prob}(j|i)=\frac{f(\lambda_{j})}{\sum_{i=1}^N f(\lambda_{i})},
	\label{eq:Pij}
\end{equation} 
where $f(\lambda)$ is an arbitrary function that represents the reputation of agents operating at rate $\lambda$ 
as seen by the population. When $f(\lambda)$ is a monotonically increasing function, the influence of agents is hierarchically organized, with fast agents having higher reputations and thus being copied more frequently, by both fast and slow agents. In this work, we use $f(\lambda)=\lambda^{\sigma}$.
\begin{figure}[t]
\centering
\includegraphics[width=\columnwidth]{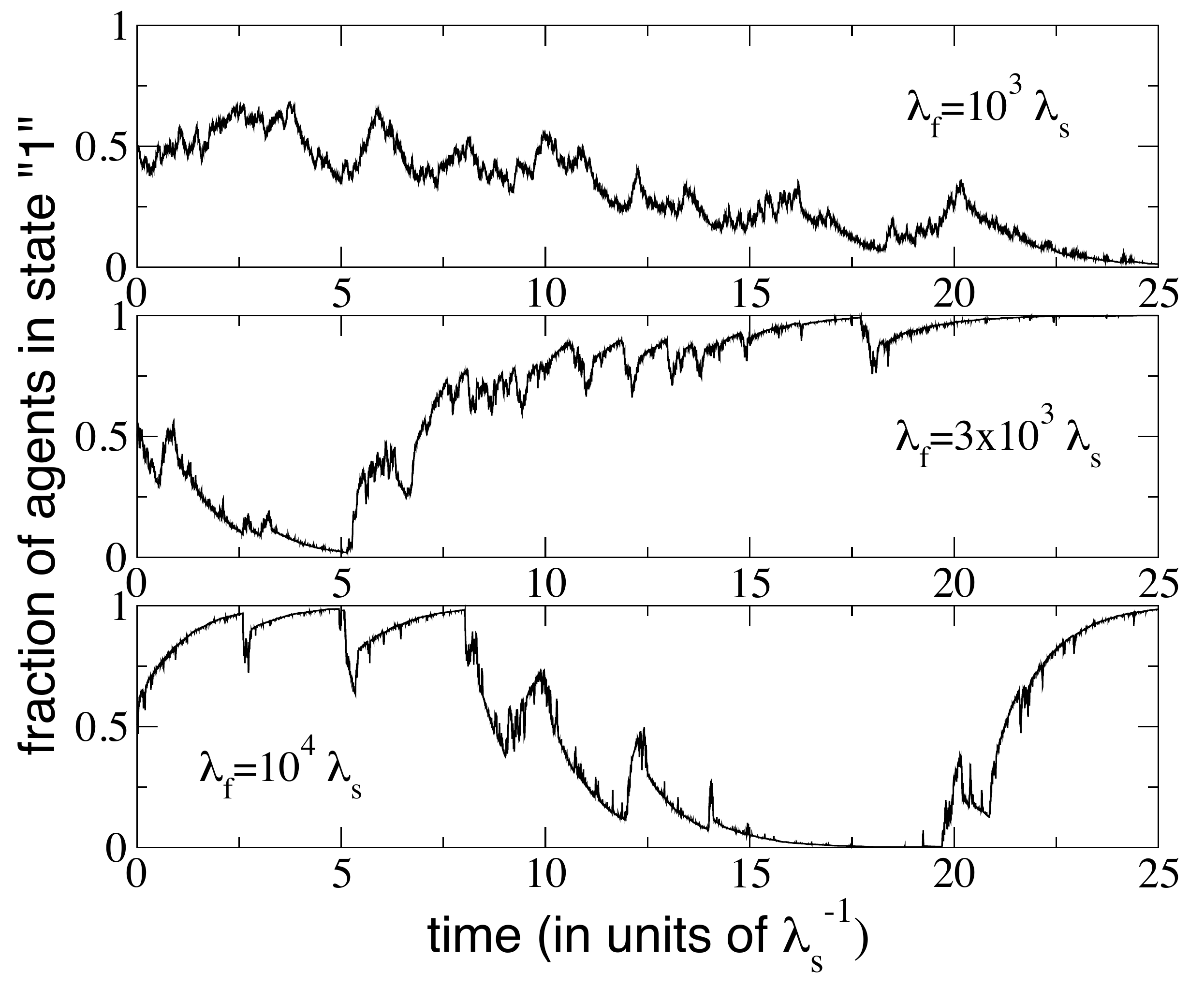}
\caption{Evolution of the fraction of agents in the ``1'' state for a two-compounded heterogeneous system with $N_f=1000$ fast and $N_s=4000$ slow agents and different time scales, $\lambda_f=10^3 \lambda_s$ (top), $\lambda_f=3 \times 10^3 \lambda_s$ (middle), and $\lambda_f=10^4 \lambda_s$ (bottom).}
\label{fig1}
\end{figure}

Figure~\ref{fig1} shows particular realizations of the process in a system consisting of a small group of fast agents, $N_f=1000$, and a large group of slow agents, $N_s=4000$. In this particular example, we set $\sigma=1$ and different time scales $\lambda_f/\lambda_s$. When the separation of the time scales between the two groups is not very important, the global dynamics is purely diffusive; as in the standard voter model (top panel of Fig.~\ref{fig1}). However, when the separation of time scales exceeds a certain critical value, the behavior changes completely. Periods of quasi-regular increase and decrease alternate, and are suddenly broken by sharp peaks. Although the system ends up in one of the two absorbing states, the peculiar pathway followed to reach consensus is not observed in the standard voter model. 
\begin{figure}[t]
\centering
\includegraphics[width=\columnwidth]{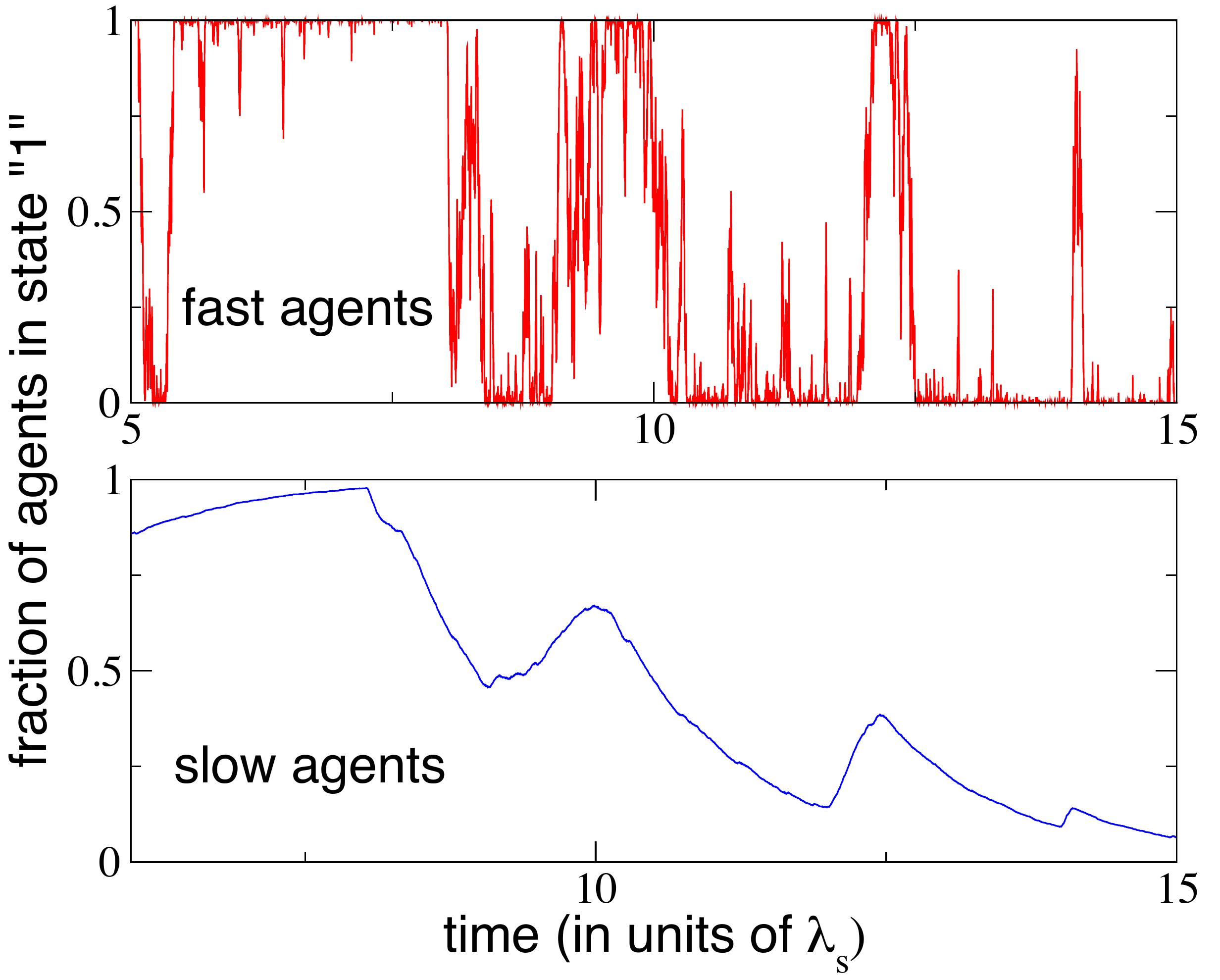}
\caption{Evolution of the fraction of fast (top) and slow (bottom) agents in state ``1'' of the same system as in Fig.~\ref{fig1}. The plots correspond to the supercritical phase with $\lambda_{f}=10^4 \lambda_s$.}
\label{fig2}
\end{figure}	

To understand this phenomenon, in Fig.~\ref{fig2} we show the temporal evolution of the two groups separately. From this figure, it is clear that the anomalous behavior we observe in Fig.~\ref{fig1} is the result of the highly differentiated dynamics of the fast and slow agents. Due to the huge differences between the time scales, from the perspective of the fast group, the slow agents seem to be frozen in their state. However, due to the monotonic increasing form of function $f(\lambda)$, the effect of slow agents on the dynamics of fast ones is small. In this situation, fast agents evolve as in the simple voter model until they reach one of their consensus states. Nonetheless, in contrast to what we observe in the simple voter model, this consensus state is not an absorbing one. Indeed, despite the low probability of a fast agent copying a slow one, the time scale of the fast agents is short enough for just this interaction to occur 
many times during the evolution of the system. When such an event occur, a fast agent may adopt an opposite opinion from a slow outsider, thus introducing some noise into the small subsystem and preventing it from becoming trapped in the consensus state. In other words, the absorbing boundary is replaced by a reflecting one. The same noise induced by slow agents can make the group of fast agents change abruptly to the opposite state, thereby providing the system with an effective two-state dynamics, as can clearly be seen in the top panel of Fig.~\ref{fig2}. 

At the same time, from the perspective of the slow group, fast agents spend long periods of time in the consensus states. Again, due to the monotonic increasing form of function $f(\lambda)$, 
slow agents have a greater tendency to copy the opinion of fast agents who, being quasi-frozen in the consensus state, act as a constant drift that pulls the opinion of the slow agents towards the opinion of the fast ones. We can interpret this as the group of slow agents becoming a herd-like group that follows the leadership of the group of fast agents. However, this behavior is not observed across the whole range of parameters and, at this point, it is unclear whether it appears suddenly at a critical value or is a crossover effect interpolating continuously from the diffusive behavior of the standard voter model to the herding behavior we observe in Fig.~\ref{fig1}.

The existence of the conserved quantity $\Phi$ implies that the dynamics cannot be completely understood only in terms of Eq.~\eqref{eq:evolucio}, as that equation does not contain any information regrading the noise in the system. So we are forced to develop a theory that includes second-order terms in the dynamics. To do this, we take advantage of the homogeneity within each group of agents and define the instantaneous average opinion state of each group as:
\begin{equation}
\Gamma_f(t) \equiv \frac{1}{N_f}\sum_{i\in fast} n_i(t) \; ; \; \Gamma_s(t) \equiv \frac{1}{N_s}\sum_{i\in slow} n_i(t).
\end{equation} 
In the limit of large systems, $\Gamma_f(t)$ and $\Gamma_s(t)$ can be considered as quasi-continuous stochastic processes in the range $[0,1]$. Furthermore, they result from the sum of a large number of random variables; so the central limit theorem can be invoked. As a result, we conclude that the stochastic evolution of the vector $\vec{\Gamma}(t)\equiv (\Gamma_f(t),\Gamma_s(t))$ can be described by a Langevin equation. In particular, for the fast group dynamics, we can write:
\begin{equation}
	\frac{d\Gamma_f(t)}{dt}=A_f\left[\vec{\Gamma}(t)\right]+\sqrt{D_f\left[\vec{\Gamma}(t)\right]}\xi_f(t),
\label{eq:langevin}
\end{equation} 
where $\xi_f(t)$ is Gaussian white noise. The drift and diffusion terms are defined respectively in terms of the infinitesimal moments as:
\begin{equation}
A_f=\frac{\langle \Delta \Gamma_f(t)|\vec{\Gamma}(t) \rangle}{dt} \; ,\;
D_f=\frac{\langle \left[\Delta \Gamma_f(t)\right]^2|\vec{\Gamma}(t) \rangle}{dt},
\label{eq:langevinterms}
\end{equation}
where $\Delta \Gamma_f(t) \equiv \Gamma_f(t+dt)-\Gamma_f(t)$~\cite{Gardiner:2004uv}. These two terms can be computed exactly using Eq.~\eqref{eq:dinamica} and read:
\begin{equation}
A_f=\alpha_{fs}(\Gamma_{s}-\Gamma_{f})
\label{eq:drift}
\end{equation}      
\begin{equation}
D_f=\frac{\alpha_{fs}}{N_f} \left( \Gamma_s+\Gamma_f \left[ 1+2 \beta_{fs}-2 \Gamma_s -2 \beta_{fs} \Gamma_f\right] \right),
\label{eq:diffusion}
\end{equation}
where we have defined: 
\begin{equation}
\alpha_{fs}=\frac{\lambda_f}{1+\beta_{fs}} \; \mbox{ and } \; \beta_{fs}=\frac{N_f f(\lambda_{f})}{N_s f(\lambda_{s})}. 
\end{equation}
Similar equations can be derived for the slow group by replacing the index $f \leftrightarrow s$ in the preceding equations.

\begin{figure}[t]
\centering
\includegraphics[width=\columnwidth]{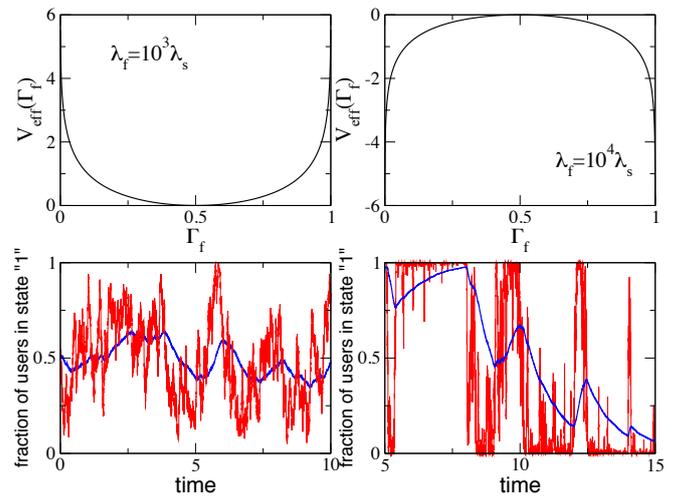}
\caption{The upper plots show the effective potential below and above the critical point in the system represented in Fig.~\ref{fig1} for a fixed value of $\Gamma_s=0.5$. The lower plots show typical realizations of the evolution of fast (red) and slow (blue) agents in both cases.}
\label{fig:FIG5}
\end{figure}

When the separation of time scales is large, the state of the slow group is perceived by the fast group as constant. In this case, we can consider $\Gamma_s$ in the previous equations as a constant parameter. As a consequence, the diffusion term in Eq.~\eqref{eq:diffusion} does not vanish when $\Gamma_f=0$, or $\Gamma_f=1$ and the system reacts at these points as it does in the presence of a reflecting barrier. Therefore, the system has a well-defined steady state controlled by an effective potential that, up to a constant value, takes the form~\cite{Gardiner:2004uv}:
\begin{equation}
V_{eff}(\Gamma_f)=\ln{D_f}-2\int \frac{A_f}{D_f}d\Gamma_f.
\end{equation}
This potential has a single extremum at approximately $\Gamma_f^*=\Gamma_s$, which changes from being a minimum to a maximum when:
\begin{equation}
2\frac{f(\lambda_f)}{f(\lambda_s)}>N_s.
\label{eq:critical}
\end{equation}
When this condition is met, the combination of a maximum at $\Gamma_f \approx \Gamma_s$ with the two reflecting barriers at $\Gamma_f=0$ and $\Gamma_f=1$ transforms the effective potential into a double-well potential with a barrier at $\Gamma_f \approx \Gamma_s$. This defines a pitchfork bifurcation that separates a diffusive phase, in which the fast group is dragged down by the slow one, from a herding phase, in which the fast group effectively behaves as a two-state system; jumping from one state to the other as in an activated process. This is illustrated in Fig.~\ref{fig:FIG5}, where we show the effective potential when $\Gamma_s=0.5$ in the two cases, along with examples of realizations of the slow and fast group dynamics. It is important to stress that condition Eq.~\eqref{eq:critical} is independent of the value of $\Gamma_s$. Therefore, even though the effective potential is modified as $\Gamma_s$ slowly evolves, its qualitative shape (whether it shows a minimum or a maximum near $\Gamma_s$) is not modified.

We should also note that, while this transition is not a true phase transition (as it disappears in the thermodynamic limit $N_s >>1$), for finite systems it behaves effectively as a first-order phase transition. Moreover, the strong separation of time scales we find in some real systems, such as speculative markets (which can be of order $\lambda_f \sim 10^{4\sim5} \lambda_s$), coupled with a growing preference function $f(\lambda)\sim \lambda^\sigma$, can result in condition Eq.~\eqref{eq:critical} holding in a quite straightforward way, even for very large populations, in particular when the exponent $\sigma>1$. 

In this paper, we present the minimal mechanisms that give rise to the emergence of leadership and herding behavior in a population of interacting agents: a strong separation of time scales coupled with some form or hierarchical organization of the influence of some agents over the others. Despite the simplicity of the toy model that we use in this work, the mechanisms are general enough to be extrapolated to more complex and realistic situations. For instance, the simple segregation of the population into only two groups is not really necessary; although mathematically more involved, it can be shown that the same phenomenology occurs in systems with a strongly heterogeneous distribution of activity rates. The hierarchical organization can also be induced by different mechanisms, such as a hierarchical organization of the network of contact among the agents, formed of a core of well-interconnected agents and a periphery of agents that are mainly connected to the core, as can be observed in many 
real complex networks~\cite{csermely:2013}. Finally, one could also argue that the influence that a group of agents has on the others is itself a stochastic process. In our case, such a scenario could easily be modeled by assigning some stochastic dynamics to the parameter $\sigma$. This is particularly interesting since, as the transition is effectively discontinuous, the dynamics would be a mixture of purely diffusive periods, during which $\sigma$ is such that the condition in Eq.~\eqref{eq:critical} is violated, and periods with strong herding behavior the rest of the time.

\begin{acknowledgments}
This work was supported by a James S. McDonnell Foundation Scholar Award in Complex Systems; the ICREA Academia prize, funded by the {\it Generalitat de Catalunya}; the MINECO project no.\ FIS2013-47282-C2-1-P; and the {\it Generalitat de Catalunya} grant no.\ 2014SGR608.
\end{acknowledgments}

%\bibliographystyle{unsrt}
%\bibliographystyle{apsrev4-1}
%\bibliography{ref_full}


\end{document}